\documentstyle[epsfig,rotating]{article}
 
\renewcommand{\baselinestretch}{1.2}

\def\etalk{{ et al., }}
\def\ie{{i.e.\ }}

\newcommand{\xbj}{\mbox{$x$}}
\newcommand{\xpom}{\mbox{$x_{I\!\!P}$}}
\newcommand {\pom} {I\hspace{-0.3em}P}
\newcommand{\naive}{\mbox{na\"{\i}ve} }

\newcommand{\ee}{\mbox{$e^+e^-$}}

\newlength{\dinwidth}
\newlength{\dinmargin}
\begin{document}

\setlength{\unitlength}{1mm}
\begin{titlepage}
\begin{flushleft}

{\tt DESY 96-014    \hfill    ISSN 0418-nnnn} \\
{\tt January 1996}                  \\
\end{flushleft}
\vspace*{3.cm}
\begin{center}
\begin{Large}
{\bf  Energy Flow in the Hadronic Final State \\
      of Diffractive and Non--Diffractive\\
      Deep--Inelastic Scattering at HERA }

\vspace{0.5cm}
{H1 Collaboration}    \\
\end{Large}
\vspace*{3.cm}
{\bf Abstract:}
\end{center}
\begin{quotation}
\renewcommand{\baselinestretch}{1.0}\large\normalsize

\noindent
An investigation of the hadronic final state in 
diffractive and non--diffractive deep--inelastic 
electron--proton scattering at HERA is presented,
where diffractive data are selected experimentally
by demanding a large gap in pseudo--rapidity around the proton remnant direction.
The transverse energy flow in the hadronic final state
is evaluated using a set of 
estimators which quantify topological properties.
Using available Monte Carlo QCD 
calculations, it is demonstrated that the final state 
in diffractive DIS exhibits the
features expected if the  interaction
is interpreted as the scattering of an electron
off a current quark with associated effects of perturbative QCD.
A model in which deep--inelastic diffraction is taken to be the
exchange of a pomeron with partonic structure is found to reproduce 
the measurements well.
Models for deep--inelastic $ep$ scattering, in which a sizeable
diffractive contribution is present because of non--perturbative effects 
in the production of the hadronic final state, reproduce the general tendencies
of the data but in all give a worse description.

\renewcommand{\baselinestretch}{1.2}\large\normalsize
 
\end{quotation}
\begin{center}
\vfill
 {\it To be submitted to Zeitschrift f.~Physik}
      \cleardoublepage
\end{center}
\end{titlepage}
%
\noindent
 S.~Aid$^{14}$,                   
 V.~Andreev$^{26}$,               
 B.~Andrieu$^{29}$,               
 R.-D.~Appuhn$^{12}$,             
 M.~Arpagaus$^{37}$,              
 A.~Babaev$^{25}$,                
 J.~B\"ahr$^{36}$,                
 J.~B\'an$^{18}$,                 
 Y.~Ban$^{28}$,                   
 P.~Baranov$^{26}$
% ...

\noindent
% ...
$ ^{12}$ DESY, Hamburg, Germany$^a$ \\
% ...
$ ^{14}$ II. Institut f\"ur Experimentalphysik, Universit\"at Hamburg,
                                     Hamburg, Germany$^ a$  \\
% ...
$ ^{18}$ Institute of Experimental Physics, Slovak Academy of
                Sciences, Ko\v{s}ice, Slovak Republic$^ f$\\
% ...
$ ^{25}$ Institute for Theoretical and Experimental Physics,
                                                 Moscow, Russia \\
 $ ^{26}$ Lebedev Physical Institute, Moscow, Russia$^ f$ \\
% ...
$ ^{28}$ LAL, Universit\'{e} de Paris-Sud, IN2P3-CNRS,
                            Orsay, France\\
 $ ^{29}$ LPNHE, Ecole Polytechnique, IN2P3-CNRS,
                             Palaiseau, France \\
% ...
$ ^{36}$ DESY, Institut f\"ur Hochenergiephysik,
                              Zeuthen, Germany$^ a$\\
 $ ^{37}$ Institut f\"ur Teilchenphysik,
          ETH, Z\"urich, Switzerland$^ i$\\
% ...
$ ^a$ Supported by the Bundesministerium f\"ur Forschung und Technologie, FRG,
        under contract numbers 6AC17P, 6AC47P, 
        6DO57I, 6HH17P, 6HH27I, 6HD17I, 6HD27I, 6KI17P, 6MP17I, and 6WT87P \\
% ...
$ ^f$ Supported by the Deutsche Forschungsgemeinschaft\\
% ...
$ ^i$ Supported by the Swiss National Science Foundation\\
\normalsize
%
\newpage
 
\section{Introduction}

 In high energy physics, the parton model together with the theory of
 strong interaction, Quantum--Chromo--Dynamics (QCD), have been shown
 to provide a good description of a variety of different processes
 involving hadrons in the final and/or initial state
 (e.g.\ in $\ee$ collisions\cite{ref:lep_qcd}, in
 hadron--hadron collisions\cite{ref:hadron_qcd} and in
 lepton--hadron scattering\cite{ref:dis_review}). 
 However, so far QCD has not been able to make predictions, derived from first
 principles, for a class of hadron--hadron collisions known
 as diffractive scattering.
 This class is 
 characterised experimentally as peripheral collisions between the 
 incoming hadrons, whereby the scattered hadrons keep their original
 identity or break up dominantly into a system of low invariant  mass. This 
 generally leads to 
 experimentally observable large rapidity gaps in these collisions.  
 Regge theory applied to hadronic interactions models diffractive 
 collisions\cite{ref:hadron_review} as the exchange 
 of an object called the pomeron ($\pom$),
 which carries only the quantum numbers of the vacuum
 and thus no colour charge.
 Attempts have been made to study the nature of 
 diffractive exchange, and it
 has been suggested that the pomeron  has partonic 
 structure\cite{ref:ingel_schlein}, indications  for
 which were found in proton--antiproton
 collisions by the UA8 experiment\cite{ref:ua8}.

At the turn--on of the electron--proton ($ep$)
collider HERA, at DESY, Hamburg, a class of 
deep--inelastic scattering (DIS) events was observed which 
exhibited in the hadronic final state an unusually large rapidity gap (LRG)
with almost no hadronic energy flow around the direction of the proton
remnant\cite{ref:zeus_lrg,ref:h1_lrg}.
Recent cross--section measurements\cite{ref:h1_f2d_93,ref:zeus_f2d_93} 
have shown that these events
exhibit characteristics similar to those of diffractive hadronic
collisions.
At low Bjorken--$x$ (see below for kinematic definitions)
 the diffractive contribution to the total DIS cross--section, and thus to the
proton structure function $F_2$, 
has been quantified in the form of a diffractive structure
function $F_2^{D(3)}$\cite{ref:h1_f2d_93,ref:zeus_f2d_93}. 
The dependence of $F_2^{D(3)}$
on the appropriate deep--inelastic kinematic variables $x_{I\!\!P}$, $Q^2$, and
$\beta$ (see below) has been measured. It demonstrates both that the diffractive
deep--inelastic $ep$ process can be interpreted as deep--inelastic
scattering of the incident electron off a colourless object
coupling to the proton in the diffractive $ep$ interaction, and that the structure of
this object is consistent with being of a partonic nature.  
 \begin{figure}[htb]
   \begin{center}
   \epsfig{file=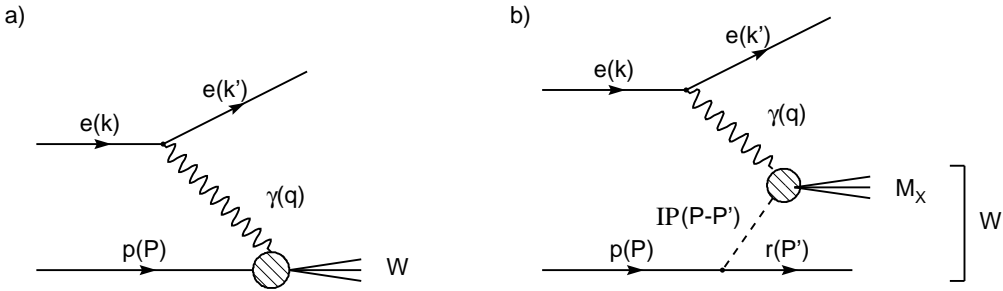,width=14.4cm}
   \end{center}
   \caption{\footnotesize Schematic picture for deep--inelastic
             scattering: non--diffractive (a) and diffractive (b).}
   \label{fig:dispict}
 \end{figure}
Thus deep--inelastic scattering may be pictured as in
figure~\ref{fig:dispict}~(a), and the diffractive contribution to it as in
figure~\ref{fig:dispict}~(b), where the process is considered in the phenomenology
which is used to quantify diffractive interactions in high energy hadron--hadron
physics.  In 
 each case a virtual photon (in the kinematic region investigated, contributions
 from $Z^0$ exchange can be neglected) probes a hadronic
 object. For non--diffractive DIS  this object is a proton, whereas in the
 diffractive case this is taken to be  a 
 colour neutral object  emitted from the proton.
The proton remnant system in the latter case
 remains colourless and thus a gap in rapidity is produced.
 Note that experimentally we observe events which
have a rapidity gap, as defined below. Throughout this paper we will however
use the terms ``diffractive'' event and ``LRG event'' synonymously.

 Using the four--momenta $k$ of the incident electron, $q$ of the 
 virtual photon and $P$ of the incident proton the following kinematical 
 variables can be defined:
  \begin{equation}
    Q^2 = - q^2,  \hspace{1cm} 
     \xbj = \frac{Q^2}{2 P \cdot q}, \hspace{1cm}  
     y = \frac{P \cdot q}{P \cdot k} \hspace{1cm} \mbox{and}  \hspace{1cm}
     W^2 = (P + q)^2,
   \label{eq:dis_var}
   \end{equation}
 where $Q^2$ is the squared virtuality of the photon and \xbj~is the Bjorken 
 scaling variable, which in the \naive Quark--Parton Model (QPM) can be interpreted
  as the fraction of the proton's momentum carried by the struck quark.
 In the rest frame of the proton $y$ is the fraction of the electron energy 
 transferred to the proton. $W$ is the  
 invariant mass of the photon--proton system and is equal to the 
 invariant mass of the total final state excluding the 
 scattered electron.
   
 Additional kinematical variables can be defined 
 using the four--momentum $P^{\prime}$ of the colourless remnant 
 (either a nucleon or a higher mass baryon excitation) in the final state:
  \begin{equation}
    \xpom = \frac{q \cdot (P - P^{\prime})}{q \cdot P},  \hspace{1.1cm}
     \beta = \frac{Q^2}{2 q \cdot (P - P^{\prime})} \hspace{1.1cm} \mbox{and} \hspace{1.1cm} 
      t = (P-P^{\prime})^2.
   \label{eq:diff_var1}
   \end{equation}
 The variables \xbj, $\xpom$~and $\beta$ are related via $\xbj = \xpom \beta$.
 Defining $M_{X}$ to be the invariant mass of the virtual 
 photon--pomeron system (i.e the invariant mass of that part of the final state
 not associated with the colourless remnant and separated from the latter by
 a LRG), $\xpom$~and $\beta$ can be written as
 \begin{equation}
    \xpom = \frac{Q^2 + M_X^2 - t}{Q^2 + W^2 - M_p^2}  
\hspace{1cm} \mbox{and} \hspace{1cm}
    \beta = \frac{Q^2}{Q^2 + M_X^2 - t},        
   \label{eq:diff_var2}
 \end{equation}
 where $M_p$ is the mass of the proton. When $M_p^2$ and $|t|$ are small
 ($M_p^2 \ll Q^2$ and $ \ll W^2$ and $|t| \ll Q^2$ and $ \ll M_X^2$),
 $\xpom$~can be  interpreted as the fraction of the proton's four--momentum 
 transferred to the pomeron, and  $\beta$  can be viewed as the fraction
 of the pomeron's four--momentum carried by the quark entering the hard 
 scattering.
 In the kinematic region under investigation both $M_p^2$ and $|t|$ can be 
 neglected and therefore $\xpom$~and $\beta$ can be calculated from $M_X^2$, $Q^2$ and
 $W^2$ as
 \begin{equation}
   \xpom  \approx \frac{M_X^2 + Q^2}{W^2 + Q^2}  \hspace{1cm} \mbox{and} \hspace{1cm}
   \beta  \approx \frac{Q^2}{M_X^2 + Q^2}.  
   \label{eq:diff_var3}
 \end{equation}

The consistency of the dependence of $F_2^{D(3)}$ on $\beta$ and $Q^2$ with
a partonic interpretation of the pomeron\cite{ref:h1_f2d_93,ref:zeus_f2d_93}
implies that the hadronic final state
in deep--inelastic diffractive scattering is 
expected to show evidence for parton production
and effects of perturbative QCD
of a nature similar to deep--inelastic $ep$ 
 scattering at appropriate ($\xbj\sim\beta$) Bjorken--$x$. 
 This picture can be further tested
 through detailed measurements of the hadronic final state. 
 In this paper we present an analysis of the hadronic
 final states for diffractive and non--diffractive DIS, and compare the
 results with QCD inspired models. 
 The analysis  concentrates
 on the observed energy flow in the laboratory frame relative to the expected direction 
 of the struck quark.  In the \naive Quark--Parton Model 
 this direction  can be calculated from the measurement 
 of the scattered electron using four--momentum conservation. 
 Such prediction of the quark direction is not
 possible in \ee  and hadron--hadron collisions. There the measured
 hadronic final state must itself be used, e.g.\ by defining
 jet directions which then can be  related to parton directions.
 In\cite{ref:zeus_eflow_93} the ZEUS collaboration compared
 the final state of events with a LRG 
 to that of events without a gap using energy flow measurements
 and concluded that in DIS with a LRG
 QCD radiation is strongly suppressed.
 Recently  a comparison of charged particle spectra for the two classes
 of events was performed by the ZEUS collaboration\cite{ref:zeus_spectra_93},
 where they observed similarities between DIS with a LRG at HERA and
 DIS at lower $W$~($\approx M_X$).
 
 In this analysis, properties of the observed 
 energy flow are defined which are sensitive to effects of perturbative QCD.
 The evolution of these properties with the kinematical variables is
 investigated. The data presented are
 corrected for detector effects. From 
 the diagrams shown in  figure~\ref{fig:dispict} 
 it is expected that the different kinematics in the two cases 
 imply differences in the 
 available phase space for QCD effects in the hadronic final state.
 The invariant mass of the system built from the photon and the probed object 
 is given by
  $W$ for non--diffractive DIS and 
 $M_X$ in the diffractive case, where $M_X \ll W$. The relevance of this
 scale  for hadron production in the diffractive process will be tested by considering
  DIS at a reduced proton beam energy. This 
 feature is also a key ingredient of several more 
 sophisticated models for  diffractive DIS, such as 
 RAPGAP\cite{ref:rapgap}. 
  Here it is assumed that 
 the hard scattering occurs off a partonic constituent of a pomeron
 emitted from the proton as shown in diagram~\ref{fig:dispict}~(b).
 QCD effects of parton showers and 
 hadronization are included in this model.
  However, alternative explanations for the diffractive process
 exist.  
 The two   models 
  LEPTO\cite{ref:lepto61} and HERWIG\cite{ref:herwig},  
 originally developed for non--diffractive DIS, 
 in the most recent versions also  generate events with a leading
 colourless remnant.
 These  models do not involve explicitly the emission of a pomeron but
 instead produce
 the gap through non--perturbative QCD effects in the evolution
 of the final state produced by deep--inelastic scattering of
 a partonic constituent of the proton.
 The predictions of these models will be compared with the data.

 The paper is organized as follows.
 After a short description of the H1 detector (section~2),
 the data taking and the event selection are briefly discussed (section~3).
 Then are described the models of the hadronic final state used in  
 deep--inelastic scattering (section~4), the procedure to correct
 for detector effects, and the sources of systematic errors (section~5).
 In section~6 the characteristic properties of the  
 energy flow are explained  and the estimators used are defined. 
 Results and the comparison with different model predictions are 
 presented in section~7. The paper is summarized in section~8.
 In an appendix the measured values of the estimators
 are listed together with their statistical and systematic errors.

\section{The H1 Detector}

 The general layout of the H1 detector 
 (described in more detail in \cite{ref:h1_detector}) is as
 follows: the interaction region is surrounded with tracking devices which
 measure the momenta of charged particles and reconstruct the 
 position of the interaction point. These tracking detectors are enclosed by a 
 calorimetric region which allows the measurement of the energy and the direction
 of charged and neutral 
 particles. All these detectors are situated within a magnetic field
 of 1.15~T, generated by a superconducting coil. The flux return yoke 
 of this coil
 is  instrumented to identify muons and to measure 
 energy escaping the main calorimeters.
 In the following the components of the H1 detector of particular relevance
 to this analysis are briefly presented. In the coordinate system
 used, the proton beam direction defines the $+z$--axis.
 The region of polar angle $0 < \theta < \pi/2$ is called ``forward region'' and
 corresponds to positive values of pseudo--rapidity $\eta=-\ln \tan \frac{\theta}{2}$,
 whereas $\eta < 0$ in the ``backward region'' ($\pi/2 < \theta < \pi$).

 The measurement of charged particle tracks and the determination
 of the interaction point is made with a system of interleaved drift and multiwire
 proportional chambers covering the central and forward regions of the
 detector ($7^o < \theta < 165^o$ corresponding to $-2.0 < \eta < 2.8$)
 over the full azimuthal range. The backward region
 is equipped with a proportional chamber (BPC) measuring charged particle
 positions in the angular range $155^o < \theta < 174.5^o$ in front 
 of the backward calorimeter (BEMC).
 
 The main calorimeter is a fine--grained liquid argon (LAr) 
 calorimeter\cite{ref:h1_lar}, covering
 a range in polar angle from $4^o$ to $155^o$ ($-1.51 < \eta < 3.35$).
 It consists of an 
 electromagnetic section with lead absorbers (20 -- 30 $X_0$) and a 
 hadronic part with 
 steel absorbers, giving a total interaction length of 5 -- 8 $\lambda$.
 The energy resolution 
 as measured in test beams\cite{ref:h1_pi}
 is $\sigma/E\approx 12\%/\sqrt E$ for 
 electrons and $\approx 50\%/\sqrt E$ for hadrons, with $E$ in 
 GeV. The energy scale is known to about $2\%$ for
 electrons and to about $5\%$ for hadrons.
 
 Covering the $\eta$ range $-3.35$ to $-1.35$ ($151^o < \theta < 176^o$), the backward
 electromagnetic calorimeter (BEMC) allows the measurement of
 the scattered electron for low $Q^2$ DIS
 events ($ 5 $~GeV$^2 < Q^2 < 100$~GeV$^2 $) 
 and provides information on hadrons scattered in this region.
 The BEMC is a  lead--scintillator sandwich calorimeter (22.5 $X_0$
 or 1 hadronic absorption length),
 with a resolution of  $\approx 10\%/\sqrt E$ for electrons and an uncertainty
 in the energy scale for electrons of about $1.7\%$.  Located behind
 the calorimeter is a time--of--flight system, consisting of 
 scintillators which veto upstream interactions of the proton beam.
 
 The very forward region is equipped with a
 copper--silicon calorimeter (PLUG), covering a range in $\eta$ from $3.54$ to $5.08$.
 The longitudinal thickness is 4.25 hadronic absorption lengths. 
 A muon spectrometer surrounds the beam pipe outside the flux return yoke
 in the forward direction. It consists of a toroidal magnet sandwiched
 between two sets of drift chambers and is used to detect charged particles
 from $ep$ interactions
 in the range $5.0 < \eta < 6.6$ by means of secondary
 particles which they produce from interactions in the beam pipe and adjunct
 material. 
 The latter two detectors are used in this 
 analysis to tag particle production in the forward region close
 to the proton beam direction.
 
 For the determination of the luminosity the process $ep \rightarrow ep\gamma$
 is used, where the electrons and photons are measured in two calorimeters
 located far downstream of the detector in the electron beam direction
 (the electron ``tagger'' at $z = -33$~m and the photon ``tagger'' at
 $z = -102$~m). 
 
\section{Data Taking and Event Selection}

 In 1993 the HERA collider operated with a 26.7~GeV electron beam and
 an 820~GeV proton beam, giving a centre of mass energy of $\sqrt{s} = $ 296~GeV.
 The data used for this analysis correspond
 to an integrated luminosity of 294~nb${}^{-1}$. The events used here were
 triggered by requiring a cluster with an energy larger than 4~GeV in the BEMC 
 and no veto from the time--of--flight system.
 This trigger has an efficiency of 99\% for
 events containing an electron with an energy of more than 10~GeV in 
 the angular acceptance of the BEMC 
 and provides a sample of DIS events at low~$Q^2$ ($ 5 $~GeV$^2 < Q^2 < 100$~GeV$^2 $).
 
  The measurement of the energy $E^{\prime}_{e}$ and
 the polar angle ${\theta}_{e}$ of the scattered electron is
 used to determine the kinematical variables 
 $Q^2 = 4 E_e E^{\prime}_{e}\cos^2({\theta}_e / 2)$ and $y = 1 - (E^{\prime}_{e} / E_e) 
 \sin^2({\theta}_e / 2)$, where $E_e$ is the electron beam energy.
 Bjorken~\xbj~and the square of the invariant mass
 of the total hadronic system, $W^2$, can be calculated from $Q^2$ and $y$
 using the relations $\xbj = Q^2 / (y s)$ and $W^2  = sy - Q^2$.
 
 The selection of deep--inelastic scattering events follows
 closely that used in the recent measurement of 
 the deep--inelastic structure function $F_2(\xbj,Q^2)$\cite{ref:h1_f2_93} and
 the diffractive structure function
 $F^{D(3)}_2(\xbj,Q^2,\xpom)$ \cite{ref:h1_f2d_93} by H1.
 The main requirements are an electron
 candidate in the backward calorimeter, a reconstructed
 interaction point, and a minimal invariant mass of the hadronic final
 state:

\begin{itemize} 
\item 
 The electron candidate is required to have an energy $E^{\prime}_{e} > 10.6$~GeV
 and to be found within the angular acceptance of the BEMC: $155^o < {\theta}_{e} < 173^o$.
 In addition, electron identification cuts to
 suppress background from photoproduction events are made
 using the information obtained from the cluster radius 
 and the matching of the cluster with a charged particle signal in the BPC.
\item
 A reconstructed vertex close to the nominal interaction point is
 demanded: $ |z_{vertex} -z^{nominal}_{vertex} | < 30$~cm. 
\item
 To ensure accurate reconstruction of the kinematics
 from the detected scattered electron, $y > 0.05$ is required since
 the reconstruction deteriorates at low values of $y$. 
\end{itemize}

 The selected events  cover the kinematical range $10^{-4} < \xbj < 10^{-2}$
 and $ 7.5 $~GeV$^2 < Q^2 < 100$~GeV$^2 $
 with an average value for $W^2$ of $23000$~GeV${}^2$ 
  ($4300$~GeV${}^2 < W^2 <53000$~GeV${}^2$).

 For the measurement of energy flow and the invariant mass of the
 final state, clusters reconstructed in the LAr calorimeter
 and in the BEMC were used.  In the BEMC, only clusters
 with an energy larger than 400~MeV are considered.

 The sample of DIS events obtained contains events with and without 
 a large gap in rapidity.  
 DIS events without a gap are selected by demanding
 \begin{itemize}
 \item
 a minimal energy deposit in the forward region measured in the LAr calorimeter:
 $E_{forward} > 0.5$~GeV (as used in
 \cite{ref:h1_eflow_92,ref:h1_bfkl}), where $E_{forward}$ is the summed
 energy in the the region $4.4^o<\theta<15^o$, corresponding to $2.03 < \eta < 3.26$.
 \end{itemize}
 This requirement together with the DIS selection described above results in
 15242 events. The only significant background
 to these events from other $ep$~interactions
 is due to photoproduction which contributes about $9\%$ of the events at
 the lowest value of \xbj~$ \approx 2 \cdot 10^{-4}$ and can be ignored for
 values of \xbj~$> 4 \cdot 10^{-4}$. 
 
 For the selection of DIS events with a large rapidity gap,
 as described in detail in \cite{ref:h1_f2d_93}, the existence of a
 region around the proton remnant direction 
 with almost no hadronic energy flow is required.
 The detector components used for this selection
 give access to the very forward region (up to a pseudo--rapidity of
 $\eta \approx 6.6$): 
 \begin{itemize}
 \item
  The energy deposited in the
 Plug calorimeter has to be smaller than 1~GeV and the number of reconstructed hit pairs
 in the forward muon spectrometer has to be smaller than 2. In addition 
 $\eta_{max,LAr} < 3.2$ is required, where $\eta_{max,LAr}$ is
 the maximum pseudo--rapidity of all clusters in the LAr calorimeter with 
 $E > 0.4$~GeV. 
 \end{itemize}
 Applying these cuts together with the DIS kinematical cuts 
 leads to a sample of 1721 events with a LRG. This sample consists of events
 where a leading hadron or cluster of hadrons $R_c$ in the forward direction
 escapes detection by remaining in the beam pipe. The acceptance in the
 invariant mass $M_R$ of $R_c$ is specified by the forward detectors used
 to define the sample. As $M_R$ increases from the mass of the proton
 to $4$~GeV the acceptance decreases from
 $100\%$ to less than $5\%$; above $4$~GeV the acceptance is
 always less than $5\%$.
 In the central detector 
 the remainder of the final state, separated by a gap from $R_c$, is detected.
 The invariant mass $M_X$ of this system defines through equation 4
 the kinematic variable $\xpom$~for each event. The data 
 sample\footnote{The sample of events with a gap and that without a 
 gap are not completely disjunct. About $5\%$ of events
 without a gap are also classified as events with a gap.
 This however does not effect the conclusions drawn.}
 used covers the range $2 \cdot 10^{-4} < \xpom < 2 \cdot 10^{-2}$.
 In all results presented, diffractive DIS is taken to be for $\xbj < \xpom < 0.02$.

\section{Monte Carlo Models for the Hadronic Final State}

The Monte Carlo models for DIS can be separated into three
parts. The hard interaction of a virtual
boson with a parton is simulated using the leading order electroweak cross--section
for parton scattering including the first order QCD correction given by the 
exact O($\alpha_S$) matrix elements (these are: the Born term for boson--quark scattering,
hard gluon Bremsstrahlung,  and the boson--gluon fusion process). 
A parton cascade includes higher order
QCD corrections to generate additional partons. The resulting coloured partons 
hadronize to give the observable particles.

The models differ mainly in the details of the parton cascade and the phenomenological 
description of the hadronization. The following description of the models emphasizes
these different approaches.

\begin{description}

\item[LEPTO]\cite{ref:lepto61} uses the
the leading--log approximation based on the Altarelli--Parisi 
evolution equations~\cite{ref:dglap}
for the parton showers.
The fragmentation is done via the Lund string 
model\cite{ref:lundstring} as implemented in JETSET\cite{ref:jetset}.

The new version (6.3) of LEPTO differs from the previous version (6.1) mainly in
two aspects. Firstly, the treatment of scattering involving a sea--quark has been
modified, motivated by the poor description of the measured transverse 
energy flow in the forward region\cite{ref:h1_bfkl} given by version 6.1.
Secondly, the possibility of colour rearrangement in the final state
through soft gluon exchange with negligible change in the momenta of
the partons has been introduced, allowing the generation of 
events with a leading colourless remnant. This soft colour interaction\cite{ref:sci}
is a non--perturbative interaction of the coloured quarks and gluons
with the colour medium
of the proton. This way of generating a colour--neutral subsystem is similar
to the ideas of Buchmueller and Hebecker\cite{ref:buchmueller}, who
performed a calculation of the diffractive cross--section on the parton
level.
The size of the diffractive contribution is determined
by a probability\footnote{The default value for
PARL(7) is 0.2, this leads to a diffractive contribution of about 7\% to
the deep--inelastic cross section in the kinematic range considered. 
A variation of this parameter between 0.1 and 0.5 gives a change in this fraction
between 5\% and 9\% but no significant changes in the measured
energy flow are observed within the two event classes.}
that such a colour exchange occurs between two 
colour charges, leading for part of the cross--section
to the formation of colour singlet subsystems separated in rapidity.

\item[ARIADNE]\cite{ref:ariadne} is a generator for QCD cascades only.
In this analysis, two versions (4.03 and 4.07) are used. The former 
version is used to calculate the correction for detector effects for 
events without a gap (see next section). Previous 
analyses\cite{ref:h1_bfkl,ref:h1_disgamm} have shown
that this version gives a reasonable description of the data.
The version 4.07 is used in comparison with  non--diffractive DIS.
For the modelling of the electroweak interaction the corresponding
parts of LEPTO\cite{ref:lepto61} are used.
Gluon radiation is performed
in ARIADNE by an implementation of the Colour Dipole Model\cite{ref:cdm}.
In this model, gluon emission from a quark--antiquark pair is treated as 
radiation from a colour dipole formed by this pair.
In contrast to the quarks in
\ee--annihilation, the proton remnant in DIS is assumed to be extended.
This leads to a suppression of the phase space for gluon radiation. 
In version 4.07 the struck quark is considered to be extended as well. The latter
modification was motivated by the disagreement between HERA data and
the ARIADNE (version 4.03) prediction in the region of the struck
quark\cite{ref:h1_bfkl,ref:h1_disgamm}.
The hadronization is done using the Lund string model (JETSET).
In version 4.07 diffractive DIS is modelled via the emission of a
pomeron from the proton and subsequent hard scattering on a partonic
constituent of the pomeron. This mechanism 
follows the spirit of RAPGAP (see below) and is not 
investigated in this paper. In\cite{ref:ariadne407} L\"onnblad demonstrates that
in the framework of the Colour Dipole Model colour reconnections cannot reproduce
the diffractive contribution to DIS in contrast to the LEPTO model.

\item[HERWIG]\cite{ref:herwig} is a general purpose generator for high energy hadronic
processes.  The 
parton shower algorithm (in the leading logarithmic approximation) takes into
account colour coherence as well as
soft gluon interference.
The hadronization in HERWIG is
performed using the concept of cluster fragmentation, where 
gluons are split non--perturbatively into 
quark--antiquark pairs. The latter are combined into colour singlet
clusters, which are split further until a minimum value of the cluster mass
is reached.
In the most recent version (5.8d)\cite{ref:herwig58d}, as used in this
analysis, HERWIG can generate a diffractive contribution to DIS,
the size of which is sensitive to a 
parameter\footnote{The chosen value for PSPLT is 0.7, this leads
to a diffractive contribution of 6\%. A variation of this
parameter between $0.5$ and $1.0$ gives contributions of 2\% and 8\%. However
the change in PSPLT results in significant changes in the predicted 
energy flow properties for both classes of events in contrast to LEPTO.}
determining the mass distribution for the cluster splitting in the 
hadronization.

\end{description}

For all the above models the MRS(H)\cite{ref:mrsh} 
set of parton distributions for the proton 
was used. These were determined using data from various
experiments including the $F_2$ measurements
from HERA made in 1992\cite{ref:h1_f2_92,ref:zeus_f2_92}.
In addition to the kinematical selection as described above
a cut on the summed energy $E_{forward}$ of all particles produced in the
range $2.03 < \eta < 3.26$ is performed 
by demanding $E_{forward} > 0.5 $~GeV for the distributions
obtained from the models in the non--diffractive case.

It should be stressed that the LEPTO model as well as the HERWIG model
in the most recent versions can generate a diffractive contribution
to the DIS cross--section compatible with the
measurement without explicitly involving the concept of  deep--inelastic
electron--pomeron scattering. These events are selected
by demanding that $\eta_{max}^{gen} < 3.2$, where $\eta_{max}^{gen}$
is the maximum pseudo--rapidity of all particles with
$E > 0.4$~GeV and  $\eta < 6.6$. Both models broadly reproduce the
$\xpom$ dependence of $F_2^{D(3)}$ as measured in\cite{ref:h1_f2d_93}
and thus the one of the deep--inelastic diffractive cross--section.
Most of the parameters in these models which
influence the description of the hadronic final state are restricted by
measurements of non--diffractive DIS, leaving at present only a few  parameters
free for the modelling of diffractive DIS. Events with a
LRG also occur in the previous versions of the models
but at a very small rate (being due to extreme fluctuations
in the fragmentation).

\begin{description}

\item[RAPGAP]\cite{ref:rapgap}
 models diffractive DIS  by
the emission of a pomeron from the proton, described by a flux factor depending only
 on  $\xpom$~and $t$. The pomeron is taken to be an object with
 a partonic structure described by parton densities,
which depend on $\beta$ and $Q^2$.
This hypothesis is consistent with the recent measurements of the
diffractive structure function $F_2^{D(3)}$\cite{ref:h1_f2d_93,ref:zeus_f2d_93}.
In the version used (1.3) the parton content
of the pomeron can be chosen to be either a quark--antiquark pair 
or two gluons. Within this analysis, the parton densities $p(z)$ are chosen to
be ``hard'' distributions ($[z\cdot p(z)] \sim z\cdot (1-z)$),
where $z$ is the fraction of the momentum of the parton
relative to the pomeron. A soft parton density
($[z\cdot p(z)] \sim (1-z)^5$) is excluded by the measurement of
$F_2^{D(3)}$ as described in \cite{ref:h1_f2d_93}.
A mixture of ``hard quark'' and ``hard gluon'' densities
was used such that an equal number of events
are generated for both densities in the kinematic range considered.
This leads to a sample in which, for $\beta \le 0.1$, more than  $70\%$
(and for $\beta \ge 0.9$ less than $5\%$) of the events are of the 
``hard gluon'' type. The mixture has been chosen to get a good
description of the measured diffractive
structure function $F_2^{D(3)}$ and was found to be able to
describe the measured energy flow.
Additional partons due to QCD radiation are generated using the Colour Dipole model,
the subsequent hadronization is performed with JETSET. 
The version of RAPGAP used does not include the evolution
of parton densities  with $Q^2$. 

\end{description}

\section{Correction for Detector Effects}
 
The correction of the measured energy flow for detector effects in 
non--diffractive DIS is done
using events generated with the ARIADNE~4.03\cite{ref:ariadne} model,
which have been passed through a simulation of the H1 detector response.
They are reconstructed in the same 
way as is done for the data. For diffractive DIS  the same procedure
is applied using the RAPGAP~1.3\cite{ref:rapgap} model.

For each distribution shown, a set of bin--by--bin correction factors is 
calculated by forming the ratio of the bin--contents for the generated events (at the
particle level) to the corresponding bin--contents for the reconstructed
events (from the detector simulation, \ie at the detector level). To 
obtain the corrected value, the raw data bin--contents have to be 
multiplied by the correction factor. The derived
factors vary only moderately from bin to bin and have values typically
between 0.8 and 1.2.

In the determination of the systematic error associated with the correction
factors, the following effects were considered:

\begin{itemize}

\item Variation of the energy scale for the scattered electron in the BEMC. The energy
   scale is known to $\pm 1.7\%$.

\item Uncertainty in the polar angle of the scattered  electron. An error
   of $\pm 2$~mrad was taken into account.

\item Uncertainty in the hadronic energy scale in the LAr calorimeter. From studies using $p_T$--balance
   between the scattered electron and the hadronic final state, the energy scale is
   known to an accuracy of $\pm 5\%$.

\item Uncertainty in the hadronic energy scale in the BEMC. The hadronic energy scale in the BEMC
 was assumed to be known to $\pm 20\%$.

\item Dependence on the model used for corrections. For events without a LRG,
  a comparison between the correction factors obtained using the ARIADNE and 
  the LEPTO model was performed. In the case of DIS with a LRG, the 
  RAPGAP model assuming either a quark parton density or 
  a gluon density alone was studied.

\item The effect of initial state photon radiation off the electron has been
  estimated with the DJANGO  model\cite{ref:django}.
  
\item The effect of background from photoproduction ($Q^2 \approx 0$) has been
      investigated using H1 photoproduction data in which the scattered 
      electron is detected in the electron calorimeter of the luminosity system
      and an electron candidate is found in the BEMC. The values
      of the estimators obtained were found to be smaller than
      those in the DIS data. Using a value of $9\%$ for the contamination 
      from photoproduction at the lowest value of 
      $\xbj \approx 2 \cdot 10^{-4}$\cite{ref:h1_f2_93}
      (contamination negligible for $\xbj > 3 \cdot 10^{-4}$) 
      an asymmetric contribution to the systematic
      error is obtained.

\end{itemize}

All contributions have been added quadratically to give a
value of the systematic error for each bin considered. 
The error bars shown contain the
statistical error (inner bars) as well as the total error (full error
bar) which has been obtained by adding statistical and systematic
errors in quadrature.

\section{Characteristic Properties of  the Energy Flow}
 
  Within the framework of the \naive Quark--Parton Model, the measurement
  of the four--momentum of the scattered electron determines the direction
  of the quark struck in the deep--inelastic scattering. Using conservation of 
  four--momentum and assuming that the partons are massless and
  the proton remnant has negligible transverse
  momentum, the polar angle of the struck quark can be calculated from the
  energy and the polar angle of the scattered electron. The  pseudo--rapidity
  ${\eta}_q$ of the struck quark can be expressed in terms
  of the kinematical variables \xbj~and $Q^2$ as:
   \begin{equation}
    {\eta}_q = \frac{1}{2} \ln \left[ \xbj \left( \frac{\xbj s}{Q^2} - 1 \right) \frac{E_p}{E_{e}} \right] 
   \end{equation}
  where $E_p$($E_e$) is the proton (electron) beam energy and
  $s$ is the square of the centre of mass energy.
  The scattered electron and the struck quark are (in the QPM) back--to--back
  in azimuth, \ie :
   \begin{equation}
     {\phi}_q = {\phi}_{e} + \pi 
   \end{equation}
  To look at deviations from these expectations, the following two variables are used:
   \begin{eqnarray}
   {\Delta\eta} = {\eta} - {\eta}_q \\
   {\Delta\phi} = {\phi} - {\phi}_{q} 
   \end{eqnarray}
  where $\eta$ ($\phi$) denotes the pseudo--rapidity (azimuthal angle) of a 
  particle or calorimetric cluster.

 \begin{figure}[htb]
   \begin{center}
   \epsfig{file=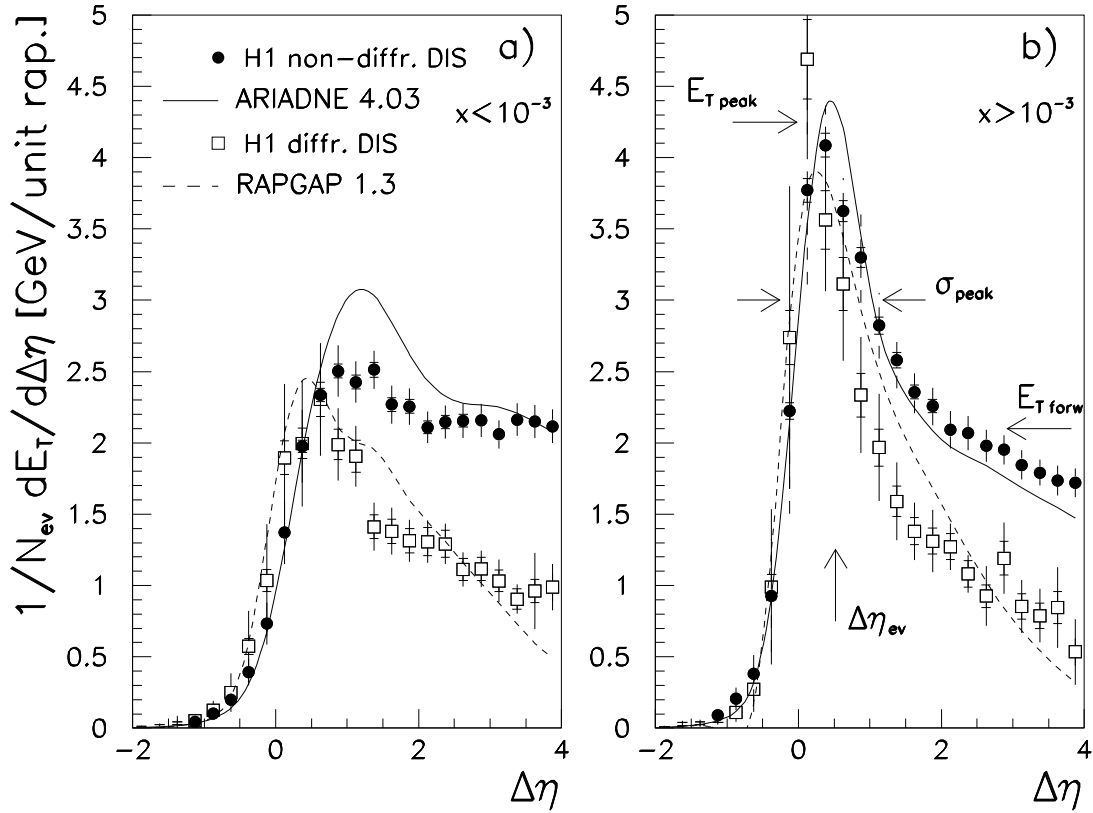,width=16.4cm}
   \end{center}
   \caption{\footnotesize Measured transverse energy flow
   relative to the calculated struck quark direction. Shown are non--diffractive
   (``non--diffr. DIS'') and diffractive  (``diffr. DIS'') data,
   both corrected for detector effects. Also
   indicated are the models for DIS (ARIADNE) and for diffractive DIS
   (RAPGAP), which have been used to do the correction.
   Data with $x < 10^{-3}$ are shown in (a), 
   those with $x > 10^{-3}$ in (b). In both figures the
   proton direction corresponds to large positive values of ${\Delta\eta}$.
   In (b) the meaning of the 4 estimators for the energy flow is also indicated
   using the non--diffractive data.}
   \label{fig:deflow}
 \end{figure}
  
  The measured transverse energy flow  $\frac{dE_T}{d\Delta\eta}$ (integrated
  over azimuthal angle) around the expected  \naive QPM direction 
  ($\Delta\eta = 0$) is shown in figure~\ref{fig:deflow}.
  Displayed are the energy flows for non--diffractive as well as for diffractive
  DIS, in two regions of \xbj~
  (\xbj~$ < 10^{-3}$ and \xbj~$ > 10^{-3}$).
   The measured transverse energy flow relative to the 
   \naive QPM prediction can be separated into that around $\Delta\eta = 0$
   (this will be denoted the ``current region''
   in the following) and the remainder of the energy flow at positive
   values of $\Delta\eta$ towards the direction of the proton remnant
   (the ``forward region'').
   The properties of the  energy flow depend strongly
   on the kinematics, as may be seen in figure~\ref{fig:deflow}. 
   The maximum of the transverse energy in the current region is shifted to 
   positive values of $\Delta\eta$ for all DIS data.
   The measured shape of the energy flows
   for the diffractive and the non--diffractive case is found to be very similar in the
   current region, whereas in the forward region  a
   reduced amount of transverse energy is expected
   for diffractive $ep$ DIS with a leading colourless remnant
   compared to the non--diffractive process.
   Also shown are the two models used to correct for detector effects,
   ARIADNE (version 4.03) for non--diffractive and RAPGAP (version 1.3)
   for diffractive DIS.

   To investigate in more detail the dependence of the hadronic final state
   on kinematical variables and to discuss the observed similarities as well
   as the differences for the two cases, 4 estimators
   of characteristic properties of the measured  transverse energy flow
   are defined as illustrated qualitatively in figure~\ref{fig:deflow}~(b).
   These estimators are calculated 
   for each event  using the measured calorimetric clusters
   and correction factors in case of the data and stable particles in the case of the
   model calculations. 
   
   Firstly, the deviation ${\Delta\eta}_{ev}$ in pseudo--rapidity of
   the maximum in transverse
   energy from the expected \naive QPM direction is determined.
   Secondly the magnitude $E_{T peak}$ of the energy flow at this position is 
   calculated. Next the width ${\sigma}_{peak}$ in pseudo--rapidity
   of the energy flow around the maximum is quantified.
   Finally the level of transverse energy $E_{T forw}$ away from
   the current region
   towards the direction of the proton remnant is determined.

   ${\Delta\eta}_{ev}$ is calculated as an $E_T$--weighted
   average of ${\Delta\eta}$ in the region with $|{\Delta\eta}| < 2$ 
   around the expected quark direction, 
   restricting the range in $\Delta\phi$
   to values with $|\Delta\phi| < 1.5$:
  \begin{equation}
        {\Delta\eta}_{ev} = \frac
           {\sum\limits_{|{\Delta\eta}|<2 ,  |{\Delta\phi}|<1.5} E_T {\Delta\eta}}
           {\sum\limits_{|{\Delta\eta}|<2 , |{\Delta\phi}|<1.5} E_T} 
   \end{equation}
   
   Having calculated the observed deviation for an event, the magnitude $E_{T peak}$ in transverse
   energy at this position (normalized to one unit of pseudo--rapidity) 
   is determined by adding up the transverse energy
   in a region of $\pm 0.25$ in pseudo--rapidity:
  \begin{equation}
         E_{T peak} = \frac{1}{0.5} \sum\limits_{|{\Delta\eta} - {\Delta\eta}_{ev}| < 0.25} E_T 
   \end{equation}
   
   The width ${\sigma}_{peak}$ in pseudo--rapidity of the energy flow in the current region is
   obtained by determining the r.m.s of the $\Delta\eta$ distribution weighted with $E_T$ 
   around the measured position ${\Delta\eta}_{ev}$ of the maximum in
   a $\Delta\eta$ range of $\pm 1$:
   \begin{equation}
        {\sigma}_{peak} = 
          \sqrt{\frac{\sum\limits_{|{\Delta\eta} - {\Delta\eta}_{ev}| < 1}
                           E_T \cdot ({\Delta\eta} - {\Delta\eta}_{ev})^2}
          {\sum\limits_{|{\Delta\eta} - {\Delta\eta}_{ev}| < 1} E_T}} 
   \end{equation}
      
   $E_{T forw}$ (normalized to one unit of pseudo--rapidity) 
   is determined in the region starting one unit of pseudo--rapidity 
   forward of the measured position of the maximum $E_T$ in the current region 
   (${\eta}_q + {\Delta\eta}_{ev}$)
   up to a fixed value of pseudo--rapidity ($\eta = 3$) -- to
   stay away from the acceptance limit of the LAr calorimeter.
   \begin{equation}
        E_{T forw} = \frac{1}{3 - ( {\eta}_q + {\Delta\eta}_{ev} + 1 )}
         \sum\limits_{( {\eta}_q + {\Delta\eta}_{ev} + 1 ) < \eta < 3} E_T  
   \end{equation}

\section{Results}
 
 In this section, the dependence of the
 measured transverse energy flow on kinematical variables
 is studied for diffractive and non--diffractive DIS, using the
 4 estimators defined in the previous section.
 First the \xbj~dependence of the estimators for diffractive and
 non--diffractive data is compared.
  Next, the non--diffractive measurements are confronted with the most
  recent versions of models for DIS.
  The characteristics of the hadronic final state in diffractive DIS are then 
  investigated for evidence of significant discrepancy from the expectation
  given by a partonic process with the related effects of perturbative QCD.
  To this end, first a phenomenological model of $ep$ DIS in the same kinematic
  region is investigated, and then the diffractive final state is compared
  with a set of different models of deep--inelastic $ep$ diffraction.
 \begin{figure}[htb]
   \begin{center}
   \epsfig{file=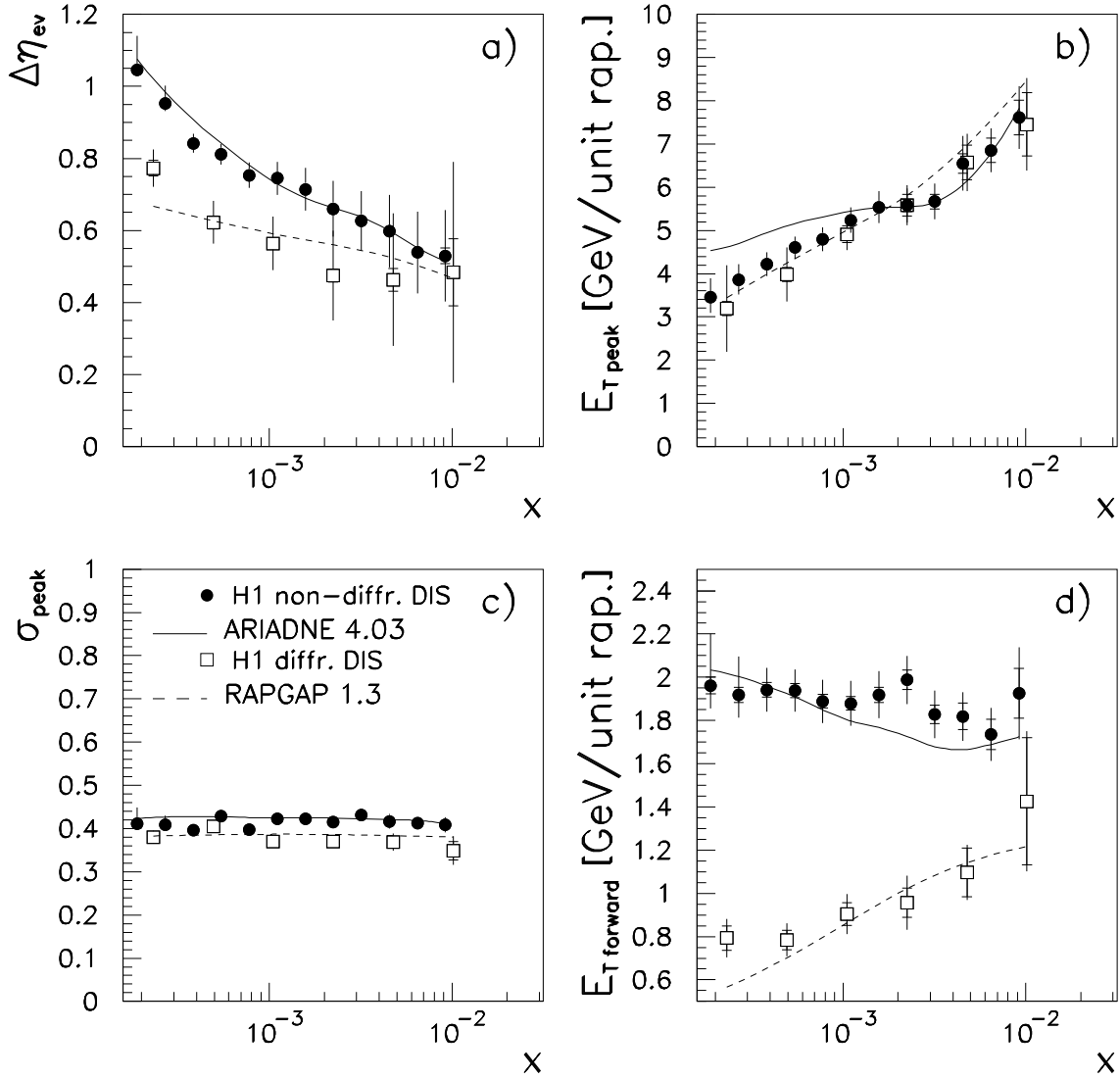,width=16.4cm} 
   \end{center}
   \caption{\footnotesize Measured estimators of the transverse energy flow for
            non--diffractive (``non--diffr. DIS'') and diffractive (``diffr. DIS'') DIS,
            compared with the ARIADNE and RAPGAP models. Shown is 
            the dependence on \xbj~for the measured deviation from the calculated
            direction (a), for the magnitude of the current region (b),
            for the width of the current region (c) and for the forward
            transverse energy (d).}
   \label{fig:deta_x}
 \end{figure}
 
 Figure~\ref{fig:deta_x} shows the dependence of the estimators on \xbj~for
 diffractive and non--diffractive DIS. 
 The measured values, together with the statistical and systematic errors,
 can be found in the appendix in tables~\ref{tab:disx} and ~\ref{tab:diffx}.
 For both cases a significant deviation $\Delta\eta_{ev}$ of
 the maximum in transverse energy from the \naive QPM expectation is observed
 which strongly increases with decreasing values of \xbj. For diffractive DIS
 the measured value is found to be smaller by about 0.2
 units in pseudo--rapidity than in the case of non--diffractive DIS.
 $E_{T peak}$ increases with
 increasing values of \xbj~with no difference visible between the two
 classes. No \xbj--dependence is observed for $\sigma_{peak}$ which has
 a value of $\approx 0.4$ (this corresponds to a full--width--half--maximum 
 for a Gaussian shaped distribution of about 0.9). Significant differences
 are observed in $E_{T forw}$,
 where for non--diffractive data $E_{T forw} \approx 2$~GeV/unit--of--rapidity
 with almost no dependence on \xbj. For the diffractive case 
 $E_{T forw}$ increases with increasing values of \xbj, the magnitude
 being lower by  20 -- 60 $\%$ compared with non--diffractive DIS,
 as expected and demonstrated clearly in the following.
 Also shown are the predictions of the  models (ARIADNE 4.03 and RAPGAP 1.3)
 that have been used to  correct for detector effects. The
 data are described reasonably well for both cases.
 At small values of \xbj~($< 10^{-3}$) the ARIADNE model (version 4.03) 
 overestimates the measured values of $E_{T peak}$.
 
 \begin{figure}[htb]
   \begin{center}
   \epsfig{file=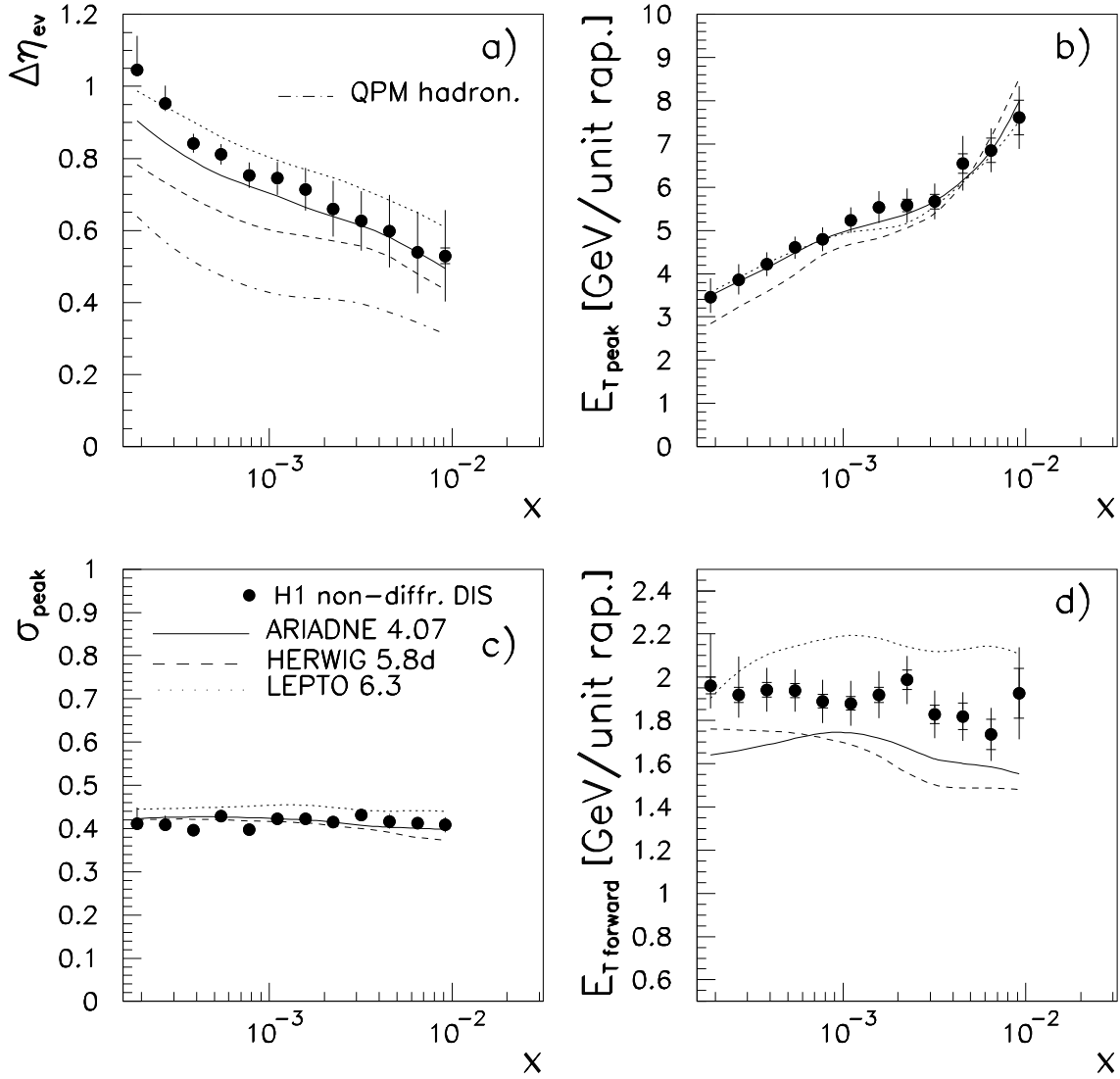,width=16.4cm}
   \end{center}
   \caption{\footnotesize Measured dependence of the estimators on \protect\xbj~
            for non--diffractive DIS, compared with the prediction
            of the ARIADNE~4.07, HERWIG~5.8d and LEPTO~6.3 models. Shown are 
             the measured deviation from the calculated
            direction (a), the magnitude of the current region (b),
            the width of the current region (c) and the forward
            transverse energy (d). In addition a QPM calculation
            including hadronization is shown in (a).}
   \label{fig:deta_xnewmodel}
 \end{figure}
 Decreasing values of \xbj~correspond on average to an increase in the invariant 
 mass $W$ of the hadronic system. This is expected to lead to an increase
 of the phase space for effects of perturbative QCD and particle production in
 the final state. The measured deviation from the \naive QPM expectation
 is sensitive to this increase and also
 to details of the implementation of QCD effects in the different models.
 In a previous analysis\cite{ref:h1_bfkl} it has been shown
 that the level of transverse energy in the central region of 
 the $\gamma^{*} p$ system (this corresponds to the forward region
 in the laboratory frame) increases with decreasing \xbj~for
 constant $Q^2$. 
 
 Before investigating in more detail the diffractive contribution, some recent 
 versions of models for DIS will be compared with non--diffractive data.
 The development of these versions of the models was motivated by the
 previously unsatisfactory description of the measured hadronic
 final state for \xbj~$ < 10^{-3}$ as shown in\cite{ref:h1_bfkl,ref:h1_disgamm}.
 The comparison of the predictions of ARIADNE (version 4.07),
 HERWIG (version 5.8d) and LEPTO (version 6.3) with the measured
 data in figure~\ref{fig:deta_xnewmodel} shows clear deviations, the differences 
 being most pronounced  in the estimators for $\Delta\eta_{ev}$ and $E_{T forw}$.
 The shape of the dependence of $\Delta\eta_{ev}$ is very
 similar for these three models but not as steep at small \xbj~as in the data.
 The predicted shape and the magnitude of the $E_{T forw}$
 dependence on \xbj~differs between models.
 The description of the data obtained with
 these models is in general worse than that given by 
 version (4.03) of ARIADNE (figure~\ref{fig:deta_x}).
 In figure~\ref{fig:deta_xnewmodel}~(a) also a QPM like calculation for the final
 state is shown.
 For this calculation (done with the LEPTO model) 
 only the contribution from the Born term for
 DIS was considered together with hadronization as implemented  in the Lund string model.
 For all values of \xbj~this calculation significantly 
 underestimates the measured deviation. It should be noted that a pure Born term 
 calculation at the parton level (i.e.\ no hadronization) gives $\Delta\eta_{ev} = 0$.
 This underlines the well known need to include effects of perturbative
 QCD (i.e.\ emission of gluons) in the modelling of the hadronic final state.
 However the understanding of the final state in the 
 new kinematic domain of DIS opened by HERA (\xbj~$ < 10^{-3}$) still remains 
 a challenge.
 
  \begin{figure}[htb]
   \begin{center}
   \epsfig{file=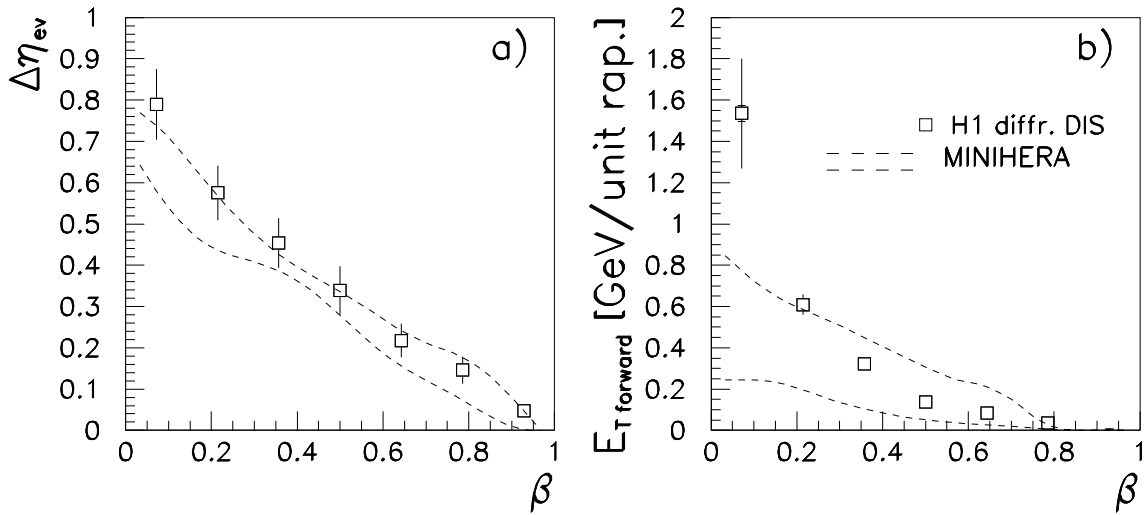,width=16.4cm}
   \end{center}
   \caption{\footnotesize Measured estimators for diffractive DIS, compared
            with a model calculation for $ep$~collisions with a reduced
            proton beam energy.
            The dashed lines indicate the range of the MINIHERA prediction
            for energies between  0.82~GeV and 8.2~GeV. Shown is 
            the dependence on $\beta$ (\xbj~for the MINIHERA model) for
            the measured deviation from the calculated
            direction (a) and for the forward
            transverse energy (b).}
   \label{fig:deta_xpombeta_minihera}
 \end{figure}
The properties of diffractive DIS  are now investigated 
in more detail. As the phase space for hadron production
depends on the invariant mass of the final state, is expected
that a large part of the differences in the $E_T$ flow observed
in the diffractive and non--diffractive case are due to the 
fact that in diffractive DIS there is an isolated
leading colourless remnant. This expectation is tested by comparing a
 model calculation (called ``MINIHERA'') for deep--inelastic scattering 
 of 26.7~GeV electrons and $f \cdot 820$~GeV protons to the diffractive data.
 For a value of $f = 0.003$ the average values of $W^2_{\mbox{\scriptsize MINIHERA}}$ 
 and $M^2_X$ (in the diffractive data) are about equal ($\approx 80$~GeV${}^2$).
 This corresponds to the average value of $\xpom$ in the data
 (\ie the average momentum of the pomeron using the
 picture of figure~\ref{fig:dispict}). To estimate the effect of 
 the spread in $\xpom$~as in the measurements, the calculation was done 
 at the values $f = 0.001$ and $f = 0.01$. For all the calculations
 the ARIADNE model (version 4.03) was used with all other
 parameters identical to those used in the calculations for
 the nominal HERA conditions. The results were cross--checked by a
 calculation with LEPTO (version 6.1) which leads to the same conclusions.
  
 Figure~\ref{fig:deta_xpombeta_minihera} shows the $\beta$ dependence of the
 two estimators $\Delta\eta_{ev}$ and $E_{T forw}$ for the diffractive data.
 Both estimators show a strong  increase with decreasing values of $\beta$
 which corresponds to $\xbj$ in DIS.
 Low values of $\beta$ 
 correspond to large masses $M_X$ of the $\gamma^{*} \pom$ system.
 The limit $\beta \rightarrow 1$
 corresponds to $M_X \rightarrow 0$ which means that the available phase
 space vanishes completely.
 The dependence of $\Delta\eta_{ev}$ on $\beta$ in the diffractive data is
 described reasonably well by the MINIHERA model calculation,
 whereas the $E_{T forw}$ dependence
 is significantly underestimated for values of $\beta < 0.2$. This
 calculation is used only to demonstrate the effect on the production of hadrons
 given the different invariant mass of the final state in the diffractive data 
 when compared to the non--diffractive data.
 The obvious limitations of the calculation 
 are the fact that the target is simply chosen to be
 a proton and furthermore a fixed value of $f$ (corresponding to $\xpom$) is
 used, in contrast to the $\xpom$ distribution in the data. However it can
 be concluded that the differences observed between the two classes of DIS data
 (figure~\ref{fig:deta_x}) can be interpreted as a result of the differences in the
 available phase space for hadron production in the final state.
 More meaningful quantitative comparisons of a model of diffraction 
 are made in the following. In this model the incident electron probes a
 pomeron with a more realistic partonic content, and therefore some of
 these shortfalls do not occur. 
 
 \begin{figure}[htb]
   \begin{center}
   \epsfig{file=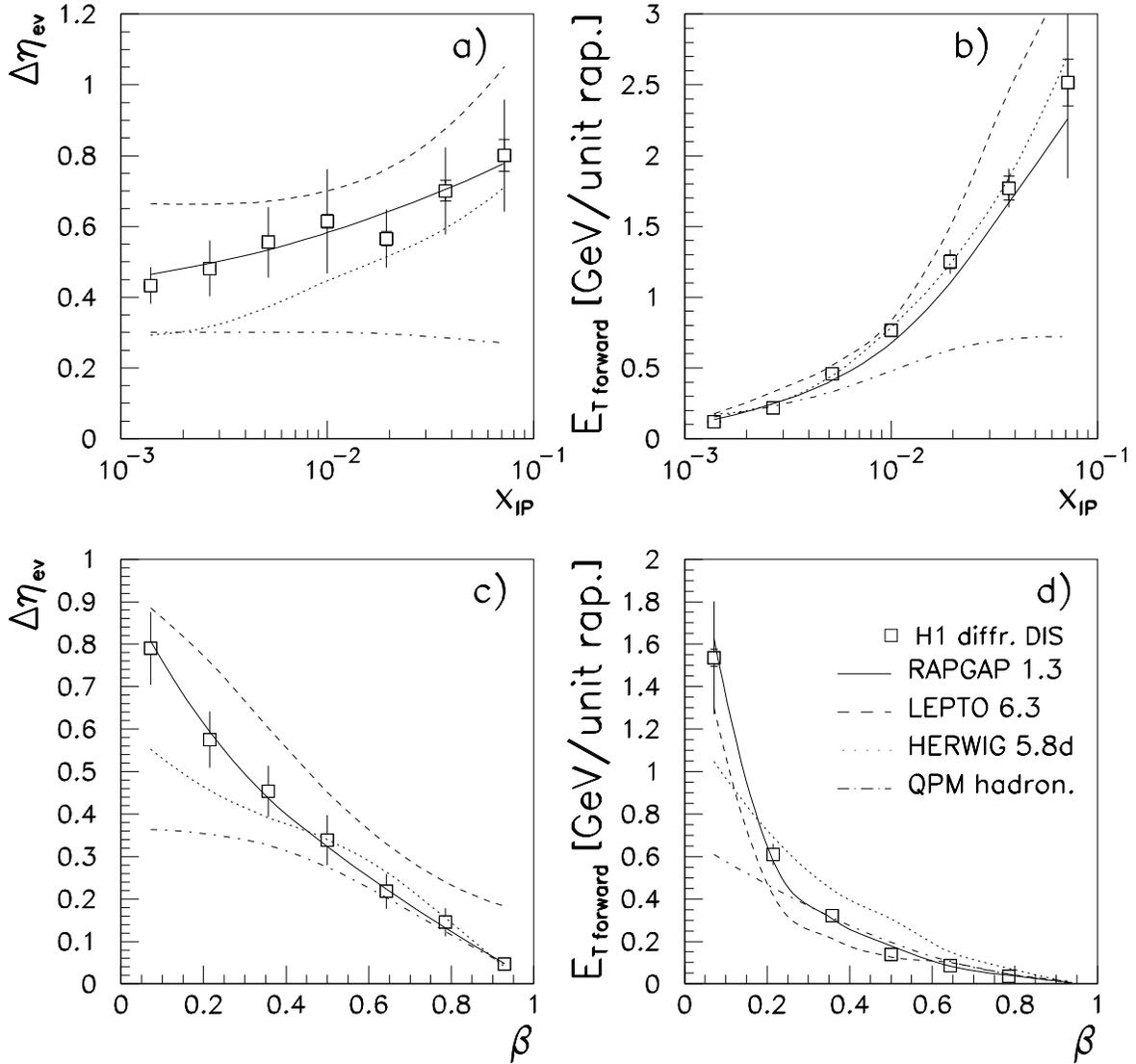,width=16.4cm}
   \end{center}
   \caption{\footnotesize Measured estimators for diffractive DIS compared
            with several different models. Shown is 
            the dependence on \protect{$\xpom$} for
            the measured deviation from the calculated
            direction (a) and for the forward
            transverse energy (b) as well as the dependence on
            \protect{$\beta$} for the deviation (c) and for the forward
            transverse energy (d). In addition a QPM calculation
            including hadronization is shown.}
   \label{fig:deta_xpombeta}
 \end{figure}
  In figure~\ref{fig:deta_xpombeta} the dependence of  ${\Delta\eta}_{ev}$ and 
 $E_{T forw}$ on  $\xpom$~is shown together with the dependence on $\beta$
 (as shown in figure~\ref{fig:deta_xpombeta_minihera})
 for diffractive DIS. The corresponding values are
 listed in tables~\ref{tab:diffxpom} and ~\ref{tab:diffbeta} in the appendix.
 The measured deviation $\Delta\eta_{ev}$ is observed to increase moderately
 whereas $E_{T forw}$ strongly increases with increasing values of $\xpom$,
 as is to be expected as the isolation of the colourless remnant in 
 diffraction from other hadrons is reduced.
 
 Also shown are the predictions of three models which can generate
 a diffractive contribution to DIS. 
 RAPGAP (version 1.3) is able to give a good description of the
 data over the whole range in $\xpom$~and $\beta$. Here diffractive DIS is
 modelled as deep--inelastic electron scattering of a partonic constituent
 of a pomeron, the latter being emitted from the proton. In the evolution of
 the final state, perturbative QCD effects are taken into account.
 A QPM like calculation is also shown.
 This calculation
 was performed with the RAPGAP model, considering only the Born term 
 (i.e.\ electron--quark scattering) and subsequent fragmentation via the
 Lund string model. This 
 calculation is observed to underestimate the data considerably in 
 kinematic regions ($\beta < 0.3$ or $\xpom > 0.01$) which 
 corresponds to larger values of $M_X$ and thus indicates the need to
 include effects of perturbative QCD  for diffractive DIS as well.
 Compared to RAPGAP the other
 two models give a worse description of the data, reproducing neither the
 shape nor the magnitude of the measurements correctly. 
 It should be noted that these discrepancies can be correlated with
 the unsatisfactory description of the non--diffractive final state 
 in these models.
 
  A previous  analysis\cite{ref:zeus_eflow_93} by the ZEUS collaboration
  comparing the energy flow 
  $\frac{dE}{d\Delta\eta}$ relative to the \naive QPM expectation
   found at most a small deviation from this
  expectation for events with a LRG and a maximum value of the deviation 
  of about 0.4 in $\Delta\eta$ for events without a gap. In the analysis presented 
  here significant deviations
  from the QPM expectation are  observed for both classes of DIS events. 
  This is due to the choice
  of $E_T$ as the weight for the pseudo--rapidity distribution
  relative to the expected QPM direction. Using $E_T$ rather than $E$
  as weighting factor leads to a greater sensitivity to deviations from the QPM
  expectation for low values of \xbj. 

  As demonstrated above, the reduction of the phase space available for
  the final state in the case of diffractive DIS is the main reason for
  the observed differences compared with the non--diffractive case.
  This conclusion was recently also reached 
  by the ZEUS collaboration in\cite{ref:zeus_spectra_93}, where
  charged particle spectra in DIS have been analyzed in the current region of the
  $\gamma^{*} p$ centre of mass system. A comparison
  of the measured transverse momenta with data from fixed target experiments
  at a value of $W$ comparable to $M_X$ in the LRG events  gave 
  good agreement.

\section{Summary and Conclusions}
 
Measurements of transverse energy flow $E_T$ in deep--inelastic
$ep$ scattering have been made using data taken at HERA with the H1 experiment.
The energy flow was analyzed in the laboratory frame of reference for
diffractive and non--diffractive data. The diffractive
data are selected experimentally by demanding 
a large rapidity gap in the hadronic final state
around the proton remnant direction, making measurements
of diffraction possible in the range $\xbj < \xpom < 0.02$.
Estimators which quantify features of the topology of the $E_T$
flow, corrected for detector effects, have been compared 
with the expectations of different models based on
QCD.

The measurements indicate that the  interpretation of 
deep--inelastic scattering as the scattering of a current quark with
associated  effects of perturbative QCD continues to be valid for
the hadronic final state of the diffractive process.
The level at which these  effects occur is consistent with 
the reduced phase space available in the diffractive process
compared to that in  non--diffractive DIS.

       The measured $E_T$ flow for diffractive DIS
       is well described by  a model (RAPGAP), in which
       the proton couples at low momentum
       transfer squared $t$ to a colourless object (pomeron).
       Here the deep--inelastic scattering process involves the
       partonic structure of the pomeron.

  Models for deep--inelastic $ep$ scattering  (LEPTO and HERWIG) in which
  the interaction of the electron involves the partonic structure of the proton,
  and not the one of an entity such as the pomeron,
  have been investigated. Here the diffractive
  configuration occurs because of non--perturbative QCD effects in the
  formation of the final state.
These models do not describe the measurements of $E_T$ flow in diffractive
DIS as well as RAPGAP. 
The observed discrepancies are however at
a level which is similar to the disagreement observed for the bulk of DIS data,
and therefore it is possible that further
developments in these models may  rectify this disagreement.

{}  \mbox{}

\noindent
{\bf Acknowledgments.} \small
 We are very grateful to the HERA machine group whose outstanding
 efforts made this experiment possible. We acknowledge the support
 of the DESY technical staff.
 We appreciate the big effort of the engineers and
 technicians who constructed and maintained the detector. We thank the
 funding agencies for financial support of this experiment. 
 We wish to thank the DESY directorate for the hospitality extended to
 the non--DESY members of the collaboration.
\normalsize 
\noindent

\begin{appendix}
\section*{Appendix}
\begin{table}[hhh]
\begin{center}
\begin{tabular}{|c||c|c|c|c|} \hline
$log_{10} \xbj$ & $\Delta\eta_{ev} [rapidity]$ & $E_{T peak} [GeV/rapidity]$ & $\sigma_{peak} [rapidity]$  & $E_{T forw} [GeV/rapidity]$  \\ \hline\hline
   -3.72 &  $ 1.05 \pm 0.01 \hspace{2mm}{}^{+\hspace{0.5mm} 0.09}_{-\hspace{0.5mm} 0.01}\hspace{0.5mm}$ &  $ 3.5 \pm 0.1 \hspace{1.8mm}{}^{+\hspace{0.5mm} 0.4}_{-\hspace{0.5mm} 0.3}\hspace{0.5mm}$ &  $ 0.41 \pm 0.01 \hspace{2.4mm}{}^{+\hspace{0.5mm} 0.04}_{-\hspace{0.5mm} 0.01}$ &  $ 1.96 \pm 0.04 \hspace{2.5mm}{}^{+\hspace{0.5mm} 0.20}_{-\hspace{0.5mm} 0.10} $ \\
   -3.57 &  $ 0.95 \pm 0.01 \hspace{2mm}{}^{+\hspace{0.5mm} 0.05}_{-\hspace{0.5mm} 0.02}\hspace{0.5mm}$ &  $ 3.9 \pm 0.1 \hspace{1.8mm}{}^{+\hspace{0.5mm} 0.4}_{-\hspace{0.5mm} 0.3}\hspace{0.5mm}$ &  $ 0.41 \pm 0.01 \hspace{2.4mm}{}^{+\hspace{0.5mm} 0.02}_{-\hspace{0.5mm} 0.01}$ &  $ 1.92 \pm 0.03 \hspace{2.5mm}{}^{+\hspace{0.5mm} 0.14}_{-\hspace{0.5mm} 0.10} $ \\
   -3.42 &  $ 0.84 \pm 0.01 \pm 0.02$ &  $ 4.2 \pm 0.1 \pm 0.3$ &  $ 0.40 \pm 0.01 \pm 0.01$ &  $ 1.94 \pm 0.03 \pm 0.10 $ \\
   -3.26 &  $ 0.81 \pm 0.01 \pm 0.03$ &  $ 4.6 \pm 0.1 \pm 0.2$ &  $ 0.43 \pm 0.01 \pm 0.01$ &  $ 1.94 \pm 0.03 \pm 0.09 $ \\
   -3.11 &  $ 0.75 \pm 0.01 \pm 0.03$ &  $ 4.8 \pm 0.1 \pm 0.3$ &  $ 0.40 \pm 0.01 \pm 0.01$ &  $ 1.89 \pm 0.03 \pm 0.10 $ \\
   -2.96 &  $ 0.75 \pm 0.01 \pm 0.05$ &  $ 5.2 \pm 0.1 \pm 0.3$ &  $ 0.42 \pm 0.01 \pm 0.01$ &  $ 1.88 \pm 0.03 \pm 0.10 $ \\
   -2.80 &  $ 0.71 \pm 0.01 \pm 0.06$ &  $ 5.5 \pm 0.1 \pm 0.3$ &  $ 0.42 \pm 0.01 \pm 0.01$ &  $ 1.92 \pm 0.04 \pm 0.10 $ \\
   -2.65 &  $ 0.66 \pm 0.01 \pm 0.08$ &  $ 5.6 \pm 0.2 \pm 0.4$ &  $ 0.42 \pm 0.01 \pm 0.01$ &  $ 1.99 \pm 0.05 \pm 0.10 $ \\
   -2.50 &  $ 0.63 \pm 0.01 \pm 0.08$ &  $ 5.7 \pm 0.2 \pm 0.4$ &  $ 0.43 \pm 0.01 \pm 0.01$ &  $ 1.83 \pm 0.04 \pm 0.10 $ \\
   -2.34 &  $ 0.60 \pm 0.01 \pm 0.10$ &  $ 6.6 \pm 0.2 \pm 0.6$ &  $ 0.42 \pm 0.01 \pm 0.02$ &  $ 1.82 \pm 0.06 \pm 0.09 $ \\
   -2.19 &  $ 0.54 \pm 0.02 \pm 0.11$ &  $ 6.9 \pm 0.3 \pm 0.4$ &  $ 0.41 \pm 0.01 \pm 0.01$ &  $ 1.74 \pm 0.07 \pm 0.10 $ \\
   -2.04 &  $ 0.53 \pm 0.02 \pm 0.13$ &  $ 7.6 \pm 0.4 \pm 0.6$ &  $ 0.41 \pm 0.01 \pm 0.02$ &  $ 1.93 \pm 0.12 \pm 0.18 $ \\  \hline
\end{tabular}
\end{center}
\caption{Values of the energy flow estimators as a function of \protect{\xbj} for
         non--diffractive DIS.
         The first error given is the statistical, the second is the systematic error.}
\label{tab:disx}
\end{table}

\begin{table}[hhh]
\begin{center}
\begin{tabular}{|c||c|c|c|c|} \hline
$log_{10} \xbj$ & $\Delta\eta_{ev} [rapidity]$ & $E_{T peak} [GeV/rapidity]$ & $\sigma_{peak} [rapidity]$  & $E_{T forw} [GeV/rapidity]$  \\ \hline\hline
   -3.64  &  $ 0.77 \pm 0.02 \pm 0.05 $ & $ 3.2 \pm0.2 \pm1.0$ &$  0.38 \pm 0.01 \pm 0.01$ & $ 0.79 \pm 0.06 \pm 0.07 $ \\ 
   -3.31  &  $ 0.62 \pm 0.02 \pm 0.06 $ & $ 4.0 \pm0.2 \pm0.6$ &$  0.41 \pm 0.01 \pm 0.01$ & $ 0.78 \pm 0.05 \pm 0.06 $ \\
   -2.98  &  $ 0.56 \pm 0.02 \pm 0.07 $ & $ 4.9 \pm0.2 \pm0.3$ &$  0.37 \pm 0.01 \pm 0.01$ & $ 0.90 \pm 0.05 \pm 0.08 $ \\
   -2.65  &  $ 0.48 \pm 0.02 \pm 0.12 $ & $ 5.6 \pm0.3 \pm0.4$ &$  0.37 \pm 0.01 \pm 0.01$ & $ 0.96 \pm 0.07 \pm 0.11 $ \\
   -2.32  &  $ 0.46 \pm 0.03 \pm 0.18 $ & $ 6.6 \pm0.4 \pm0.5$ &$  0.37 \pm 0.01 \pm 0.02$ & $ 1.10 \pm 0.11 \pm 0.06 $ \\
   -1.99  &  $ 0.48 \pm 0.09 \pm 0.29 $ & $ 7.5 \pm0.7 \pm0.8$ &$  0.35 \pm 0.02 \pm 0.03$ & $ 1.43 \pm 0.30 \pm 0.13 $ \\ \hline
\end{tabular}
\end{center}
\caption{Values of the energy flow estimators as a function of \protect{\xbj} for
         diffractive DIS.
         The first error given is the statistical, the second is the systematic error.}
\label{tab:diffx}
\end{table}
\begin{table}[hhh]
\begin{center}
\begin{tabular}{|c||c|c|} \hline
$log_{10} \xpom$ & $\Delta\eta_{ev} [rapidity]$ & $E_{T forw} [GeV/rapidity]$  \\ \hline\hline
   -2.86  &  $ 0.43 \pm 0.01 \pm 0.05 $ & $  0.12 \pm 0.01 \pm 0.02 $ \\
   -2.57  &  $ 0.48 \pm 0.01 \pm 0.08 $ & $  0.22 \pm 0.01 \pm 0.02 $ \\
   -2.29  &  $ 0.56 \pm 0.02 \pm 0.10 $ & $  0.46 \pm 0.02 \pm 0.03 $ \\
   -2.00  &  $ 0.61 \pm 0.02 \pm 0.15 $ & $  0.77 \pm 0.03 \pm 0.04 $ \\
   -1.71  &  $ 0.57 \pm 0.02 \pm 0.08 $ & $ 1.25 \pm 0.05 \pm 0.07 $ \\
   -1.43  &  $ 0.70 \pm 0.03 \pm 0.12 $ & $ 1.77 \pm 0.09 \pm 0.11 $ \\
   -1.14  &  $ 0.80 \pm 0.05 \pm 0.15 $ & $ 2.52 \pm 0.17 \pm 0.66 $ \\ \hline
\end{tabular}
\end{center}
\caption{Values of the energy flow estimators as a function of \protect{\xpom} for
         diffractive DIS.
         The first error given is the statistical, the second is the systematic error.}
\label{tab:diffxpom}
\end{table}
\begin{table}[hhh]
\begin{center}
\begin{tabular}{|c||c|c|} \hline
$ \beta$ & $\Delta\eta_{ev} [rapidity]$ & $E_{T forw} [GeV/rapidity]$  \\ \hline\hline
0.07  &  $ 0.79 \pm 0.01 \pm 0.09 $ & $1.54 \pm 0.04 \pm 0.26 $ \\
0.21  &  $ 0.58 \pm 0.01 \pm 0.07 $ & $ 0.61 \pm 0.02 \pm 0.05 $ \\
0.36  &  $ 0.45 \pm 0.01 \pm 0.06 $ & $ 0.32 \pm 0.01 \pm 0.01 $ \\
0.50  &  $ 0.34 \pm 0.01 \pm 0.06 $ & $ 0.14 \pm 0.01 \pm 0.01 $ \\
0.64  &  $ 0.23 \pm 0.01 \pm 0.04 $ & $ 0.09 \pm 0.01 \pm 0.01 $ \\
0.79  &  $ 0.15 \pm 0.01 \pm 0.03 $ & $ 0.04 \pm 0.01 \pm 0.01 $ \\
0.93  &  $ 0.05 \pm 0.01 \pm 0.01 $ & $ 0.01 \pm 0.01 \pm 0.01 $ \\ \hline
\end{tabular}
\end{center}
\caption{Values of the energy flow estimators as a function of \protect{$\beta$} for
         diffractive DIS.
         The first error given is the statistical, the second is the systematic error.}
\label{tab:diffbeta}
\end{table}
\end{appendix}

\clearpage

\renewcommand{\baselinestretch}{1.0}
{\Large\normalsize}

\end{document}